\documentclass[aps,pre,amsmath,twocolumn,amssymb,floatfixng,showpacs,longbibliography,
superscriptaddress,nofootinbib]{revtex4-2}
\usepackage{amsfonts}
\usepackage{amsmath}
\usepackage{amssymb}
\usepackage{latexsym}
\usepackage{array}
\newcolumntype{P}[1]{>{\centering\arraybackslash}p{#1}}
\newcolumntype{M}[1]{>{\centering\arraybackslash}m{#1}}
\usepackage{caption}
\usepackage[colorlinks,citecolor=red,urlcolor=blue,bookmarks=false,hypertexnames=true]{hyperref} 
\usepackage{xcolor}
\usepackage{bm}
\usepackage{relsize}
\usepackage{float}
\usepackage{subfloat}
\usepackage{dcolumn}
\usepackage{epsfig}
\usepackage{graphicx}
\usepackage{multirow}
\usepackage{makecell}
\usepackage{mathtools}
\usepackage[bbgreekl]{mathbbol}

\usepackage{footnote}

\begin{document}

\title{Quantum Brownian Motion of a charged oscillator in a magnetic field coupled to a heat bath through momentum variables}
\author{Suraka Bhattacharjee}
 \affiliation{Raman Research Institute, Bangalore-560080, India} 
\author{Urbashi Satpathi}
\affiliation{Department of Chemistry, Ben-Gurion University of the Negev, Beer-Sheva 84105, Israel}

 \author{Supurna Sinha}
 \affiliation{Raman Research Institute, Bangalore-560080, India}

\date{\today}

\begin{abstract}
We study the Quantum Brownian motion of a charged particle moving in a harmonic potential in the presence of an uniform external magnetic field and linearly coupled to an Ohmic bath through momentum variables. We analyse the growth of the mean 
square displacement of the particle in the classical high temperature domain and in the quantum low temperature domain 
dominated by zero point fluctuations. We also analyse 
the Position Response Function and the long time tails of various correlation functions. We notice some distinctive features, different from the usual case of a charged quantum Brownian particle in a magnetic field and linearly coupled to an Ohmic bath via position variables. 

\end{abstract}

\maketitle
\section{Introduction}
The Quantum Brownian Motion of a particle undergoing diffusion driven
by quantum fluctuations has been the focus of study for some time \cite{grabertone,graberttwo,decosup,Aslangul}.
More recently, researchers have investigated 
the diffusion behaviour of a charged particle in a magnetic 
field \cite{ford,Urbashione,surakaone}. To study this problem, one considers a Quantum
Langevin Equation which corresponds to a reduced description of the system in which the coupling with the
heat bath is described by two terms: an operator valued random force $F(t)$ with mean zero and a mean force
characterized by a memory function $\mu(t)$ \cite{Li}. Moreover, some of the studies have focused on an anomalous form of these couplings within a similar framework of the Quantum Langevin dynamics \cite{fa, Mckinley,Didier}. In \cite{Lisy,Tothova}, the authors have considered the motion of a harmonically oscillating Brownian particle in the presence of a bath that is affected by a harmonic potential. It has been reported that the fluctuation dissipation theorem still remains valid at equilibrium, however the random forces show dependence on the confining potential. Moreover, the effect of the harmonic potential on the dissipation constant has been determined in cold atom experiments and also in Molecular Dynamics (MD) simulation for a molecular solute particle suspended in a fluid medium \cite{Berner,Daldrop}. Although, in most of the literature, one considers a conventional form of coupling between the 
particle and the bath via position variables \cite{grabertone,graberttwo,decosup,Aslangul,ford,Urbashione,surakaone}, several authors have also considered the case of such a quantum Brownian particle coupled to the bath variables 
through momentum coordinates \cite{Caldeira,Leggett,Cuccoli,Bai,Bao,Bao2,Ankerhold,Malaymomentum}. 

In \cite{Malaymomentum} the authors have derived a Quantum Langevin Equation of a charged quantum particle moving in a harmonic potential in the presence of an uniform external magnetic field and linearly coupled to a quantum heat bath through momentum variables. Here we use the Quantum Langevin Equation derived in \cite{Malaymomentum} as a starting point to analyse the growth of the mean 
square displacement of the particle in the classical high temperature domain and in the quantum low temperature domain 
dominated by zero point fluctuations. Furthermore, we analyse 
the long time tails of various correlation functions. In addition to it, we have shown that the random forces in this case are dependent on the confining potential, although we have not considered the effect of the harmonic potential on the bath variables explicitly in the Hamiltonian \cite{Lisy,Tothova}. The modified force correlations were derived in \cite{Malaymomentum} for the case of a system coupled to a bath via momentum variables and the fluctuation dissipation theorem holds at equilibrium.

The outline of the paper is as follows. In Sec $II$ we 
discuss the Quantum Langevin Formalism and the position correlation function,
the mean square displacement, the Position Response Function, the position-velocity 
correlation function and the velocity autocorrelation
function. In Sec $III$ we discuss the long time behaviour 
of the various correlation functions mentioned in Sec $II$.
In Sec $IV$ we present our results and discussions. 
Finally we end the paper with some concluding remarks in 
Sec $V$. We present details of some of the calculations
in the Appendix.

\section{Quantum Langevin formalism }
The Hamiltonian of a charged Brownian particle in the presence of a magnetic field and a 
harmonic potential and linearly coupled to a heat bath consisting of $N$ oscillators, 
via momentum coordinate, is given by \cite{Malaymomentum}:
\begin{align}
    H=&\frac{1}{2m}\left(\vec{p}-\frac{q}{c}\vec{A} \right)^2+\frac{1}{2}m {\omega}_0^2 r^2+ \notag\\ 
    &\sum_{j=1}^N\left[\frac{1}{2m_j}\left(\vec{p}_j-g_j \vec{p}+ \frac{g_j q}{c}\vec{A} \right)^2+\frac{1}{2}m_j {\omega}_j^2 q_j^2 \right]
\label{QLE}
\end{align}
where, $ m $ is the mass of the Brownian particle, $\vec{p}$ and $\vec{r}$ are the momentum and position coordinates of the particle respectively. ${\omega}_0$ is the frequency of the harmonic oscillator potential, $ q $ is the charge of the particle, $ c $ is the speed of light and $ \vec{A} $ is the vector potential pertaining to the magnetic field $\vec{B}$. $m_j$, $\vec{p}_j$, $\vec{q}_j$, ${\omega}_j$ denote the heat bath variables corresponding to the mass, momentum, position and frequency of the $j^{th}$ oscillator respectively. ${g_j}$ is the coupling strength between the particle and the $j^{th}$ bath oscillator.\\
In Eq.(\ref{QLE}) the first term pertains to the kinetic energy in the presence of the external magnetic field. The second term pertains to the harmonic oscillator 
potential and the last term consists of the Hamiltonian of interaction between the charged particle and the bath of oscillators and the bath Hamiltonian.
The first step in deriving the Quantum Langevin equation (QLE) of a charged Brownian particle is to write the Heisenberg equations of motion for both the Brownian particle variables and the bath variables.  
Solving the Heisenberg equations of motion for the bath variables $\vec{q}_j,\vec{p}_j$ and substituting the solutions for the bath variables in the equation of motion of the Brownian particle $\vec{r},\vec{p}$, one can arrive at the Quantum Langevin Equation (QLE) of a quantum particle of charge $q$ moving in a harmonic potential in the presence of an uniform external magnetic field and linearly coupled to a quantum heat bath through momentum variables \cite{Malaymomentum}:\\
The Quantum Langevin Equation (QLE) of a Brownian particle of charge $q$ moving in a harmonic potential in the presence of an uniform external magnetic field $\vec{B}$ and linearly coupled to a quantum heat bath through momentum variables \cite{Malaymomentum}:
\begin{align}
\notag m_r \ddot{\vec{r}}(t)+\int\mu (t-t')\dot{\vec{r}}(t')dt'+m_r {\omega}_0^2 \vec{r}(t)+\mu_d (t)\vec{r}(0)-\\
\frac{q}{c}(\dot{\vec{r}}(t)\times\vec{B})=\vec{F}(t) \notag \label{Langevineq}\\
\end{align}
where, $\vec{r}(0)$ is the initial position of the particle at $t=0$, $m_r$ is the normalized mass of the particle given by:
\begin{align}
    m_r=m/\left[1+\sum_{j=1}^N\frac{g_j^2 m}{m_j}\right]
\end{align}
with ${g_j}$ being the coupling strength between the particle and the $j-th$ bath oscillator. \\
 $ \vec{F}(t) $ is the random force given by,

\begin{align}
\vec{F}(t)=&\sum_{j=1}^N g_j m_r \omega_{j}^{2}\vec{q}_{j}^{h}(t)\Theta(t)
\end{align} 
where, $\Theta(t)$ is the step function and $\vec{q}_j^h$ is the solution of the homogeneous equation of motion for the heat bath variables given by:
\begin{align}
    q_j^h(t)\equiv q_j(0)cos(\omega_j t)+\frac{p_j(0)}{m_j \omega_j}sin(\omega_j t)
\end{align}
q$_j$(0) and p$_j$(0) are the initial position and momentum of the j$^{th}$ bath particle respectively \cite{Malaymomentum}.\\
$\vec{F}(t)$ has the following properties:
\begin{align}
\langle F_{\alpha}(t)\rangle=&0\\
\notag \frac{1}{2}\langle{\lbrace F_{\alpha}(t),F_{\beta}(0)}\rbrace\rangle =&\frac{\hbar \delta_{\alpha \beta}}{ 2\pi}\int_{-\infty}^\infty{{d{\omega}}Re[\mu_d({\omega})]}
\frac{{\omega}^3 m_r}{{\omega}_0^2 m}  \\ \hspace{0.2cm} \smaller{\times}  \coth(\frac{\hbar{\omega}}{2k_BT}) e^{-i{\omega} t} \label{noisecorA}\\
\langle{[F_{\alpha}(t),F_{\beta}(0)}]\rangle=&\frac{\hbar\delta_{\alpha \beta}}{ \pi}\int_{-\infty}^\infty{{d{\omega}}Re[\mu_d({\omega})]}
\frac{{\omega}^3 m_r}{{\omega}_0^2 m} e^{-i{\omega} t}\label{forcecorrelationA}
\end{align}

$\mu(t)=\mu_{d}(t)+\Gamma\mu_{od}(t)$ is the memory kernel,
$ \mu_d $ and $\mu_{od}$ are the diagonal and off-diagonal parts of the memory kernel $\mu$, 
{\begin{align}
\mu_{d}(t)=&\sum_{j=1}^{N}\frac{g_{j}^2 m m_r \omega_{0}^2}{m_j}\cos(\omega_j t)\Theta(t)\\
\mu_{od}(t)=&\sum_{j=1}^{N}\frac{g_{j}^2 m m_r \omega_{j}\omega_c}{m_j B}\sin(\omega_j t)\Theta(t)\\
\end{align}
\begin{align}
 \Gamma=&\begin{pmatrix}
    0 & B_z & -B_y\\
    -B_z & 0 & B_x\\
    B_y & -B_x & 0
    \end{pmatrix}
\end{align}
$B_x , B_y, B_z$ are the components and $B$ is the magnitude of the uniform magnetic field $\vec{B}$.
Notice that the diagonal elements of the memory kernel have an explicit dependence on the external harmonic oscillator 
potential via $\omega_0$. The Fluctuation Dissipation Theorem 
(Eq.(\ref{noisecorA}) and Eq.(\ref{forcecorrelationA}) connects the force correlation to a memory kernel with a diagonal element dependent on the harmonic 
potential via $\omega_0$ \cite{Tothova}. 
There are several papers in the literature where
a similar dependence of the memory kernel on the externally applied potential has been studied \cite{Berner,Daldrop,Lisy,Tothova}.}

\hspace*{0.2cm}Now, one can define an effective random force $\vec{G}(t)$ as follows: 
$\vec{G}(t)=\vec{F}(t)-\mu_d(t)\vec{r}(0)$, where $\vec{G}(t)$ retains all the statistical properties of $\vec{F}(t)$ \cite{Li,Malaymomentum}.\\
\hspace*{0.2cm}For the Ohmic model, $\mu_d(t)=2\gamma m \delta(t)$ \cite{Urbashione,surakaone}. \\

Using this, Eq.(\ref{Langevineq}) takes the form:
\begin{align}
 m_r \ddot{\vec{r}}(t)+m \gamma\dot{\vec{r}}(t)+m_r \omega_0^2 \vec{r}-
\frac{q}{c}(\dot{\vec{r}}(t)\times\vec{B})=\vec{G}(t) \label{Langevineq2}
\end{align}

Taking the Fourier transform of Eq.(\ref{Langevineq2}) and separating it into $x$ and $y$ components we get:

\begin{align}
    -m_r\omega^2x+im\gamma\omega x+m_r \omega_0^2 x-im \omega_c \omega y=\mathcal{G}_x(\omega) \label{langevinomega1}\\
     -m_r\omega^2y+im\gamma\omega y+m_r \omega_0^2 y+im \omega_c \omega x=\mathcal{G}_y(\omega) \label{langevinomega2}
\end{align}
 where, $\omega_c=\frac{qB}{mc}$ and $\mathcal{G}_x(\omega)$ and $\mathcal{G}_y(\omega)$ are the Fourier transforms of $G_x(t)$ and $G_y(t)$ respectively.\\
 
 Solving Eqs.(\ref{langevinomega1}) and (\ref{langevinomega2}), we get $x(\omega)$ and $y(\omega)$ as:
 \begin{widetext}

 \begin{align}
x(\omega)=-\frac{-(m_r \omega_0^2-m_r \omega^2+i m \omega \gamma)(-\mathcal{G}_x)-i \omega \omega_c m(-\mathcal{G}_y)}{m^2\omega_c^2\omega^2-(m_r \omega_0^2-m_r\omega^2+i m\omega \gamma)^2} \label{xomega}\\
y(\omega)=-\frac{-(m_r \omega_0^2-m_r \omega^2+i m \omega \gamma)(-\mathcal{G}_y)+i \omega \omega_c m(-\mathcal{G}_x)}{m^2\omega_c^2\omega^2-(m_r \omega_0^2-m_r\omega^2+i m\omega \gamma)^2}\label{yomega}
 \end{align}
 \end{widetext}
 
 The position autocorrelation function as defined in Refs.\cite{surakaone,surakatwo} is given by:
 \begin{align}
     C_x(\omega)=\frac{1}{2}\langle \lbrace x(\omega),x^* (\omega)\rbrace \rangle \label{Cxanticommut}\\
     C_y(\omega)=\frac{1}{2}\langle \lbrace y(\omega),y^* (\omega)\rbrace \rangle \label{Cyanticommut}
 \end{align}
 
 We have made use of the force correlation functions mentioned in the Appendix A (see Eq.(\ref{noisecorA})), Eq.(\ref{xomega}-\ref{yomega}) and Eq. (\ref{Cxanticommut}-\ref{Cyanticommut}) to get $C_x(\omega)=C_y(\omega)=C(\omega)$ as:\\
 
 \begin{align}
    &C(\omega)= \notag \\&\frac{ \left(\frac{\hbar\gamma m_r}{ \omega_0^2}  \right) \omega^3 \left[\frac{1}{m^2}(\omega_0^2-\omega^2)^2 +\frac{1}{m_r^2}\omega^2(\omega_c^2+\gamma^2)\right] coth \left( \frac{\omega}{\Omega_{th}}\right) }{\left[\left \lbrace \frac{m_r}{m}\left(\omega_0^2-\omega^2 \right)^2+\frac{m}{m_r} \omega^2 \left(\omega_c^2+ \gamma^2 \right)   \right \rbrace^2-4 \omega^2 \omega_c^2 \left(\omega_0^2-\omega^2  \right)^2 \right]} \label{Comegageneral}
\end{align}
 
 $\Omega_{th}=\frac{2 k_B T}{\hbar}$\\

According to the Fluctuation Dissipation Theorem (FDT) in the Fourier domain \cite{Felderhof1,Weber,surakaone,surakatwo}:
\begin{align}
 \frac{1}{\hbar} C(\omega)= coth \left(\frac{\omega}{\Omega_{th}}\right) R''(\omega)  \label{flucdiss} 
\end{align}
where R$''(\omega$) is the imaginary part of the Response Function, which is defined as the measure of the response of a system to an external perturbing force. In the time domain, the displacement of a particle in response to a perturbing force can be expressed as the convolution of the response function and the applied force.\\
In this paper, we study the position autocorrelation function, mean square displacement, response function, position-velocity correlation function and velocity autocorrelation function in the context of a quantum 
Brownian motion of a charged oscillator in a magnetic field coupled to a bath via momentum coupling.\\

\textbf{(i)} The \textbf{Mean Square Displacement (M.S.D)} is defined by \cite{Urbashione}:
\begin{align}
    M.S.D= 2(C(0)-C(t))
\end{align}

We can express C(t) in terms of its Fourier transform C($\omega$) as follows:
\begin{align}
    C(t)=&\frac{1}{2\pi}\int_{-\infty}^{\infty} C(\omega) e^{-i \omega t} d\omega \notag\\
    =&\frac{\hbar}{2\pi}\int_{-\infty}^{\infty} R''(\omega)  coth \left(\frac{\omega}{\Omega_{th}}\right) e^{-i \omega t} d\omega \label{Ccorr}
\end{align}
\\
R$''(\omega$) has the following poles in the lower half plane:\\ \\
p$_1=\frac{-m \omega_c -i m \gamma +\sqrt{4 m_r^2 \omega_0^2+m^2(\omega_c + i \gamma)^2}}{2 m_r}$,\\ p$_2=-\frac{m \omega_c +i m \gamma +\sqrt{4 m_r^2 \omega_0^2+m^2(\omega_c + i \gamma)^2}}{2 m_r}$, \\ p$_3=\frac{m \omega_c -i m \gamma +\sqrt{4 m_r^2 \omega_0^2+m^2(\omega_c - i \gamma)^2}}{2 m_r}$, \\ p$_4=\frac{-m \omega_c +i m \gamma +\sqrt{4 m_r^2 \omega_0^2+m^2(\omega_c - i \gamma)^2}}{2 m_r}$ \\
and $coth\left(\frac{\omega}{\Omega_{th}} \right)$ has poles at $(-in\pi \Omega_{th})$, $n$ being any positive integer.\\
 So, using the Cauchy's Residue Theorem, the position correlation function C(t) can be calculated by summing over the residues at the poles \cite{graberttwo,Urbashitwo,surakatwo}:
  \begin{align}
 C(t)=C_1(t)+C_2(t)
 \end{align}
 where,
\begin{align}
&C_1(t)=(- i)\sum_{j=1}^4 \left( Res[C(\omega)exp(-i\omega t), p_j] \right) \\ 
&C_2(t)=  \left(-i \hbar \Omega_{th} \right) \sum_{n=1}^\infty R''(-i n \pi \Omega_{th}) exp(-n \pi \Omega_{th}t) \label{Ccorrtwo}
 \end{align}
 $Res[C(\omega) exp(-i\omega t), p_j]$ pertains to the residue of C($\omega$) corresponding to the pole at p$_j$.\\ \\
 \textbf{(ii)} The \textbf{Position Response Function (R(t))} is calculated from the position autocorrelation, by the application of the Kramers Kronig transformation and the Fluctuation dissipation theorem Eq.(\ref{flucdiss}).
 As already discussed, the imaginary part of the Position Response Function in the frequency domain can be derived from the correlation function using the Fluctuation dissipation theorem Eq.(\ref{flucdiss}). So, from Eq.(\ref{Comegageneral}) we get R$''$($\omega$) as:

\begin{align}
    &R''(\omega)= \notag \\&\frac{\left(\frac{\gamma m_r}{ \omega_0^2}  \right) \omega^3 \left[\frac{1}{m^2}(\omega_0^2-\omega^2)^2 +\frac{1}{m_r^2}\omega^2(\omega_c^2+\gamma^2)\right] }{\left[\left \lbrace \frac{m_r}{m}\left(\omega_0^2-\omega^2 \right)^2+\frac{m}{m_r} \omega^2 \left(\omega_c^2+ \gamma^2 \right)   \right \rbrace^2-4 \omega^2 \omega_c^2 \left(\omega_0^2-\omega^2  \right)^2 \right]} \label{imresponse}
\end{align}

The real part ($R'(\omega)$) of the Position Response Function can be calculated from the imaginary part ($R''(\omega)$) using the Kramers Kronig relation given by \cite{bohren2010did,surakaone}:
\begin{align}
    R'(\omega)= \frac{1}{\pi}\int_{-\infty}^{\infty}\frac{\omega' R''(\omega')}{(\omega'^2-\omega^2)}d\omega' \label{Kramerskronig}
\end{align}
 $R'(\omega')$ has poles at $\omega'$=p$_j$ (for j=1,...4) and at $\omega'=\pm \omega$.\\
The residues at $\omega$ and $-\omega$ cancel out. Thus $R'(\omega)$ is derived using the Cauchy's residue theorem:
\begin{align}
  R'(\omega)=(-2 \pi i)\sum_{j=1}^{4}(Res[\frac{\omega' R''(\omega')}{\pi(\omega'^2-\omega^2)},p_j])  \label{}
\end{align}

The total Position Response Function is then calculated as:
\begin{align}
    R(\omega)=R'(\omega)+iR''(\omega) \label{Romega}
\end{align}
Then the Position Response Function R(t) in the time domain is easily derived via the Fourier transform of R($\omega$) obtained from the above formalism:
\begin{align}
    R(t)=\frac{1}{2\pi}\int_{-\infty}^{\infty}R(\omega)e^{-i \omega t}d\omega
\end{align}
This calculation shows that the response function exhibits a decay from a finite value at t=0, which is distinct from the position coupling case which shows a rise in the response function starting from t=0 and settling to a constant as t increases. \\
For the simplest case, one may consider the particle in the absence of a magnetic field, i.e, $\omega_c$=0 and the response function works out to be:
\begin{align}
 R(t)=&\left[\frac{\left \lbrace \left(\frac{m}{m_r} \gamma \omega_p \right) cos\left(\frac{\omega_p t}{2} \right)+ \left(\omega_p^2-2 \omega_o^2 \right)sin\left(\frac{\omega_p t}{2} \right) \right \rbrace}{  m \omega_0^2\omega_p}\right] \notag \\ &  \times exp(-\frac{m \gamma t}{2 m_r}) \label{responsenomag}
\end{align}
where, $\omega_p=\sqrt{4 \omega_0^2-\frac{m^2}{m_r^2} \gamma^2}$. \\

It can be seen from Eq.(\ref{responsenomag}) that, at t=0, R(t) reduces to $\frac{\gamma}{ m_r \omega_0^2}$ and we notice a subsequent exponential decay at long times, in the over-damped regime. However, it can be seen that in the presence of a magnetic field ($\omega_c \neq$0), the decay is followed by an oscillatory behaviour when $\omega_c> \gamma$. \\ Interestingly, this form of the Position Response Function 
is similar to that of the Velocity Response Function that one gets in the case 
of a coupling to the bath via position variables. It could perhaps be due to the fact that the momentum and the position are related via a time derivative and
this leads to the difference that we observe between the two cases (the case of a bath coupled via position variables and the one in which the bath is coupled
via momentum variables). \\

\textbf{(iii)} The \textbf{Position-Velocity Correlation Function (C$_{x-v}$(t))} is derived from the position autocorrelation function \cite{surakaone}:
\begin{align}
    C_{p-v}(t)=\frac{d}{dt} C(t) \label{Cx-v(t)}
\end{align}
In the Fourier space:
\begin{align}
     C_{p-v}(\omega)=-i \omega C(\omega) \label{Cx-v(omega)}
\end{align}
Using Eqs.(\Ref{Comegageneral}), (\Ref{Cx-v(t)}) and (\Ref{Cx-v(omega)}):
\begin{align}
    C_{p-v}(t)= C_{p-v_1}(t)+ C_{p-v_2}(t)
\end{align}
where,
\begin{align}
    &C_{p-v_1}(t)= (- i)\sum_{j=1}^4 \left( Res[C_{p-v}(\omega) exp(-i \omega t), p_j] \right) \\
  &C_{p-v_2}(t)=\left(i \hbar \pi \Omega_{th}^2 \right) \sum_{n=1}^\infty n R''(-i n \pi \Omega_{th}) exp(-n \pi \Omega_{th}t)
\end{align}

\textbf{(iv)} Similar to the case of the position-velocity correlation function, the \textbf{Velocity Autocorrelation Function (C$_v$)} is calculated as follows \cite{surakaone}:
\begin{align}
    C_{v}(t)=-\frac{d^2}{dt^2} C(t) \label{Cv(t)}
\end{align}
In the Fourier space:
\begin{align}
     C_{v}(\omega)=\omega^2 C(\omega) \label{Cv(omega)}
\end{align}
Using Eqs.(\Ref{Comegageneral}), (\Ref{Cv(t)}) and (\Ref{Cv(omega)}):
\begin{align}
    C_{v}(t)= C_{v_1}(t)+C_{v_2}(t)
\end{align}
where,
\begin{align}
   & C_{v_1}(t)= (- i)\sum_{j=1}^4 \left( Res[C_{v}(\omega) exp(-i\omega t), p_j] \right) \\
 & C_{v_2}(t)=\left(i \hbar \pi^2 \Omega_{th}^3 \right) \sum_{n=1}^\infty n^2 R''(-i n \pi \Omega_{th}) exp(-n \pi \Omega_{th}t) 
\end{align}

\section{Long time behaviour of the correlation functions}

In this section, we have analyzed the long time behaviour of the correlation functions at finite low temperatures and extended the results to get the behaviour in the zero temperature Quantum regime. 
\subsection{Position Autocorrelation Function}
As discussed in the previous section, the position autocorrelation function at any temperature is given by Eq.(\Ref{Ccorr}). At low temperatures, exp(-$\gamma$t) falls off faster than  exp(-n$\pi \Omega_{th}$t) and the long time behaviour of the correlation function is determined by C$_2$(t) (see Appendix for details).
Therefore, at finite low temperatures at long times we get:
\begin{align}
    C(t)=C_2(t)= \left(-i \hbar \Omega_{th} \right) \sum_{n=1}^\infty R''(-i n \pi \Omega_{th}) exp(-n \pi \Omega_{th}t) \label{Cfinitetemp}
\end{align}

Now, expanding the imaginary part of the response function R$''(\omega)$ in a power series using Eq.(\ref{Comegageneral}) and Eq.(\ref{flucdiss}),  we get:
 \begin{align}
     &R''(\omega)=\frac{\gamma}{ m_r \omega_0^6}\omega^3+ 
     \frac{(2 m_r^2 \omega_0^2-m^2(\gamma^2-3 \omega_c^2))\gamma}{ m_r^3 \omega_0^{10}}\omega^5+\notag\\ &\frac{(3m_r^4 \omega_0^4-4m^2m_r^2 \omega_0^2(\gamma^2-3 \omega_c^2)+m^4(5\omega_c^4-10\omega_c^2\gamma^2+\gamma^4))}{m_r^6 \omega_0^{12}} \notag \\
     &\times\left(\frac{\gamma m_r}{ \omega_0^2}  \right) \omega^7+........\label{Comega}
 \end{align}
Using the above power series expansion, we obtain from Eq.(\ref{Cfinitetemp}) \cite{graberttwo,surakatwo}:
 \begin{align}
   & C(t)=-i(\hbar \Omega_{th})\big[\frac{\gamma}{ m_r \omega_0^6}(-i\pi \Omega_{th})^3 \sum_{n=1}^\infty n^3 exp(-n \pi \Omega_{th}t)+\notag\\
     &\frac{\gamma (2 m_r^2 \omega_0^2-m^2(\gamma^2-3\omega_c^2))}{m_r^3 \omega_0^{10}}(-i\pi \Omega_{th})^5\sum_{n=1}^\infty n^5 exp(-n \pi \Omega_{th}t)\notag \\
    &+ .....\big]
     \end{align}
     \begin{align}
& C(t)  = \frac{\hbar \pi^3 \Omega_{th}^4 \gamma}{ m_r \omega_0^6} \frac{e^{\pi \Omega_{th}t}(1+4e^{\pi \Omega_{th}t}+e^{2\pi \Omega_{th}t})}{(e^{\pi \Omega_{th}t}-1)^4}- \notag\\
  & \frac{\hbar \pi^5 \Omega_{th}^6 \gamma(2 m_r^2 \omega_0^2-m^2(\gamma^2-3 \omega_c^2))}{ m_r^3 \omega_0^{10}}\times \notag \\
 & \frac{e^{\pi \Omega_{th}t}(1+26e^{\pi \Omega_{th}t}+66e^{2\pi \Omega_{th}t}+26e^{3\pi \Omega_{th}t}+e^{4\pi \Omega_{th}t})}{(e^{\pi \Omega_{th}t}-1)^6} \notag \\
 &+.....
     \end{align}
At finite temperatures and long times, ${\pi \Omega_{th}t}>>1$ and thus we get an exponential decay of the position autocorrelation function:
\begin{align}
    C(t)=A_cexp[-\pi \Omega_{th}t] \label{poshightemp}
\end{align}
where,
\begin{align}
    &A_c=\frac{\hbar \pi^3 \Omega_{th}^4 \gamma}{ m_r \omega_0^6}-\frac{\hbar \pi^5 \Omega_{th}^6 \gamma(2 m_r^2 \omega_0^2-m^2(\gamma^2-3 \omega_c^2))}{ m_r^3 \omega_0^{10}} \notag \\
    &+......
\end{align}

In contrast, at T$\longrightarrow0$, we expand $e^{\pi \Omega_{th}t}$ as a Taylor series and we get a power law behaviour, similar to the case of a particle coupled to a bath via a position coordinate coupling \cite{graberttwo,surakatwo}, albeit, with higher power exponents.

 \begin{align}
 C(t)=A_{1c} t^{-4}+A_{2c}t^{-6}+...\label{poslowtemp}
 \end{align}
 where,
 \begin{align}
& A_{1c}=\frac{6 \hbar \gamma}{2\pi m_r \omega_0^6}, \notag\\
 &A_{2c}=\frac{-120 \hbar \gamma(2 m_r^2\omega_0^2-m^2(\gamma^2-3\omega_c^2))}{2\pi m_r^3 \omega_0^{10}}. \notag
 \end{align}
 It is evident from Eq.(\ref{poslowtemp}) that at long times the position autocorreation function falls off as t$^{-4}$, which is faster than the t$^{-2}$ fall off of the position autocorrelation function of a particle coupled to the heat bath via a position coordinate coupling \cite{surakatwo}.

 \subsection{Position-Velocity Correlation Function}
 
 The position-velocity correlation function is calculated using Eq.(\ref{Cx-v(t)}) and the long time behaviour is determined at low temperatures.\\
 At finite temperatures, the position velocity correlation function is derived from Eq.(\ref{poshightemp}):
 \begin{align}
     C_{p-v}(t)&=-\pi \Omega_{th} A_{c}(t)exp[-\pi \Omega_{th}t] \notag\\
     &=A_{c_{p-v}}(t)exp[-\pi \Omega_{th}t]
 \end{align}
 where $A_{c_{p-v}}(t) = -\pi \Omega_{th} A_{c}(t)$.\\ 
 At T$\rightarrow0$, 
 \begin{align}
     C_{p-v}(t)=A_{1c_{p-v}}t^{-5}+A_{2c_{p-v}}t^{-7}+....
 \end{align}
 where,
 \begin{align}
 &A_{1c_{p-v}}=-4A_{1c},\notag\\
 &A_{2c_{p-v}}=-6A_{2c} \notag
 \end{align}
 \subsection{Velocity Autocorrelation Function}
 
 The velocity autocorrelation function is calculated using Eq.(\ref{Cv(t)}) and the long time behaviour is determined at low temperatures.\\
 At finite temperatures, the velocity autocorrelation function is derived from Eq.(\ref{poshightemp}):
 \begin{align}
     C_{v}(t)=&-\pi^2 \Omega_{th}^2 A_{c}(t)exp[-\pi \Omega_{th}t]\notag\\
     =&A_{c_{v}}(t)exp[-\pi \Omega_{th}t]
 \end{align}
 where, $A_{c_{v}}(t)=-\pi^2 \Omega_{th}^2 A_{c}(t)$.\\ 
 At T$\rightarrow0$, 
 \begin{align}
     C_{v}(t)=A_{1c_{v}}t^{-6}+A_{2c_{v}}t^{-8}+....
 \end{align}
 where,
 \begin{align}
& A_{1c_{v}}=5A_{1c_{p-v}}=-20A_{1c},\notag\\
 &A_{2c_{v}}=7A_{2c_{p-v}}=-42A_{2c} \notag
 \end{align}
 
 \section{Results and Discussions}
 In this section, we have plotted and analyzed our results based on the graphical representations of the derived quantities.\\
 We have used dimensionless variables in our graphical representations. In Fig.(1) we have plotted the scaled total response function $R(t)/R(0)$ as a function of scaled time $\omega_0 t$ for under-damped and over-damped cases. Figs.(2) and (3) display the mean square displacement (${MSD}$), scaled by the magnetic length squared (r$^2_B={\frac{\hbar c}{eB}}$)  and the scaled correlation functions at a finite temperature and very low temperature (T$\rightarrow$0) respectively for various damping regimes.
 
 \begin{widetext}
 
\begin{figure}[H]
\centering
\includegraphics[scale=0.9]{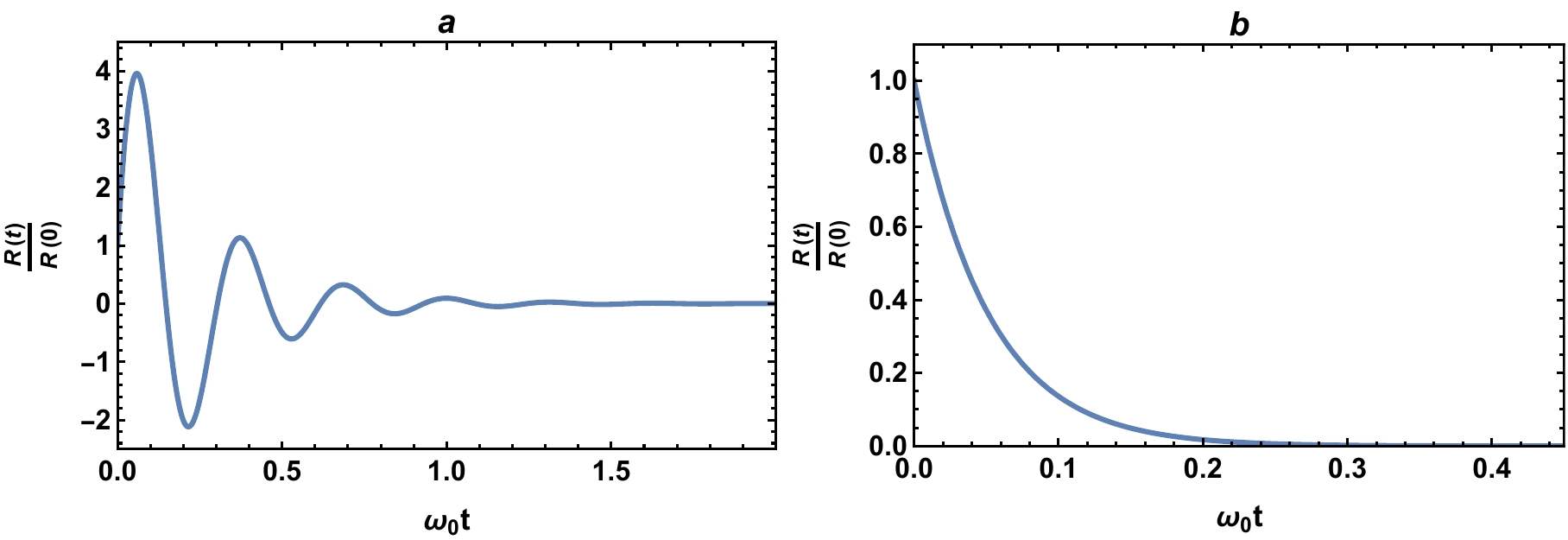}

\caption{The scaled response function $\left(\frac{R(t)}{R(0)} \right)$ versus time for ${\omega}_0=1$, m=1, m$_r$=0.5: (a) ${\omega}_c$=10, ${\gamma}=2$; (b) ${\omega}_c$=1.2, ${\gamma}=10$ }
\end{figure}

\end{widetext}
\begin{widetext}

\begin{figure}[H]
\centering
\includegraphics[scale=0.82]{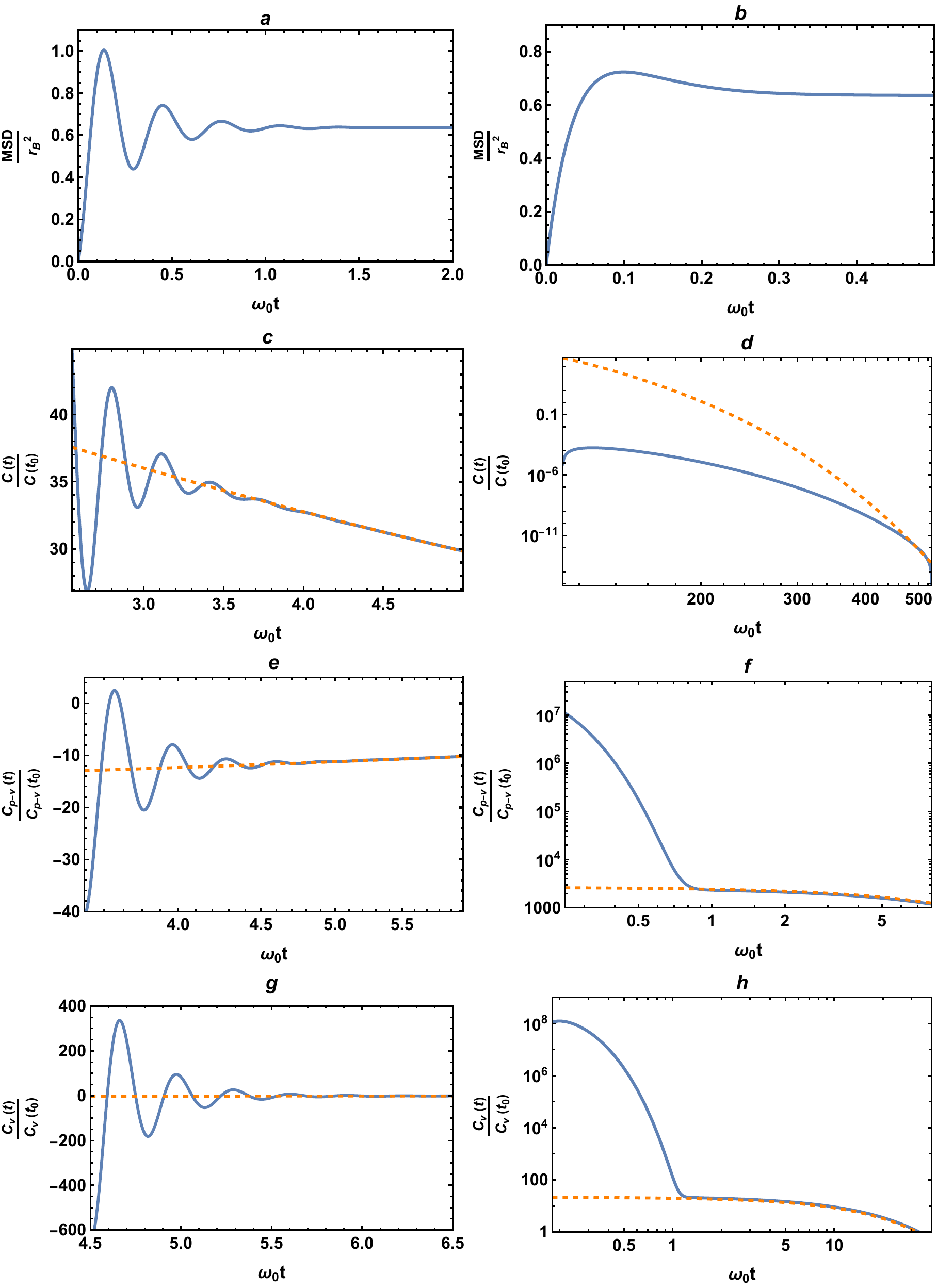} 
\caption{The scaled mean square displacement $\left(\frac{MSD}{r^2_B}\right)$  and scaled correlation function plots as a function of time at a finite temperature for ${\omega}_0=1$, m=1, m$_r$=0.5, ${\Omega}_{th}$=0.03: (a) $\frac{MSD}{r^2_B}$ for ${\omega}_c$=10, ${\gamma}=2$; (b) $\frac{MSD}{r^2_B}$ for ${\omega}_c$=1.2, ${\gamma}=10$; (c)  $\frac{C(t)}{C(t_0)}$ for ${\omega}_c$=10, ${\gamma}=2$; (d) $\frac{C(t)}{C(t_0)}$ for ${\omega}_c$=1.2, ${\gamma}=10$  in log-log scale; (e) $\frac{C_{p-v}(t)}{C_{p-v}(t_0)}$ for ${\omega}_c$=10, ${\gamma}=2$; (f)$\frac{C_{p-v}(t)}{C_{p-v}(t_0)}$ for ${\omega}_c$=1.2, ${\gamma}=10$  in log-log scale; (g) $\frac{C_v(t)}{C_v(t_0)}$ for ${\omega}_c$=10, ${\gamma}=2$; (h) $\frac{C_v(t)}{C_v(t_0)}$ for ${\omega}_c$=1.2, ${\gamma}=10$  in log-log scale (The scaling time $t_0=\frac{1}{\Omega_{th}}$).[The orange dashed lines in the plots represent the long time exponential behaviours of the correlation functions].}
\end{figure}

\begin{figure}[H]
\centering
\includegraphics[scale=0.82]{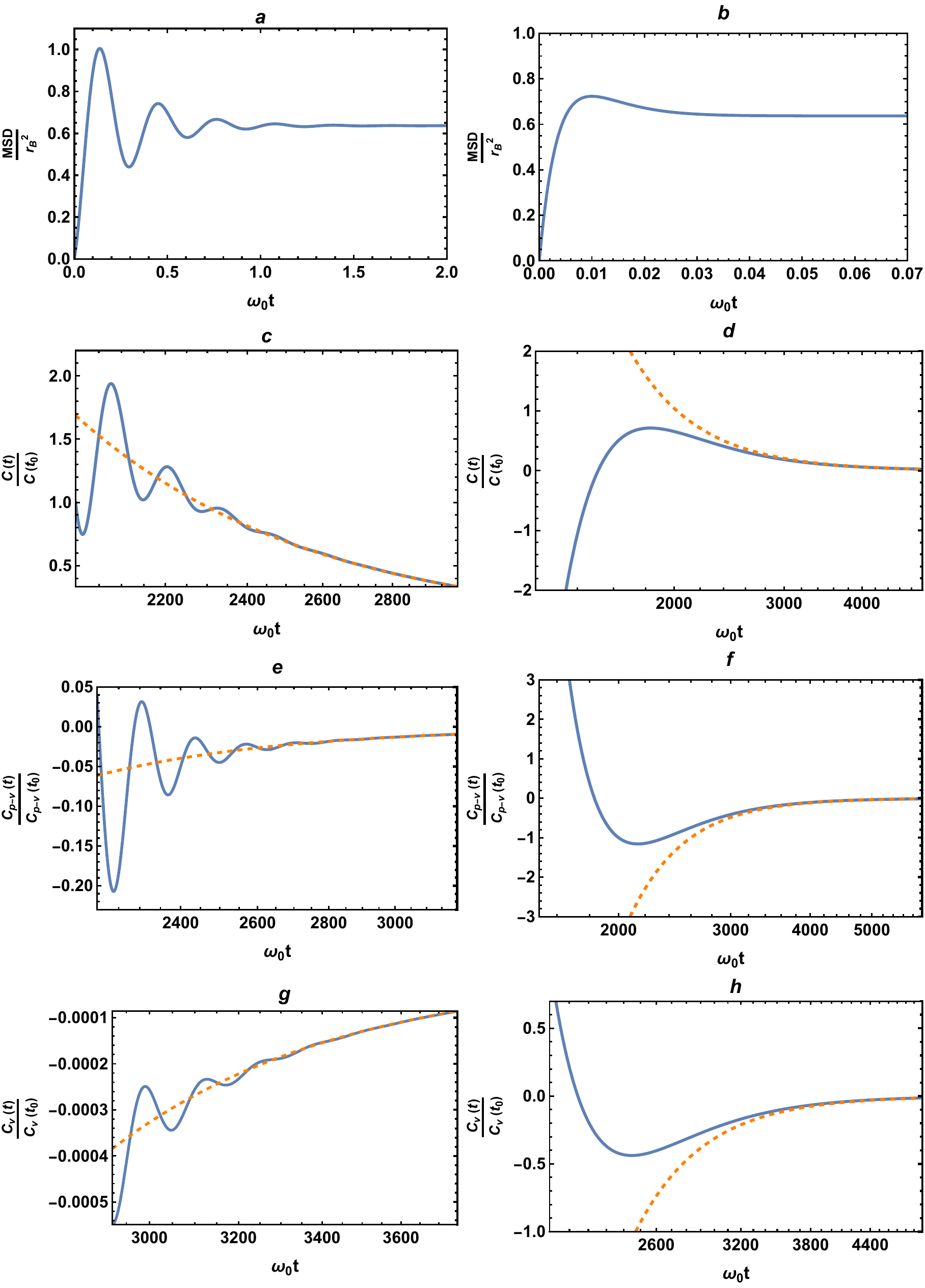} 
\caption{The scaled mean square displacement $\left(\frac{MSD}{r^2_B}\right)$ and scaled correlation function plots as a function of time at very low temperature for ${\omega}_0=1$, m=1, m$_r$=0.5, ${\Omega}_{th}$=0.0005: (a) $\frac{MSD}{r^2_B}$ for ${\omega}_c$=10, ${\gamma}=2$; (b) $\frac{MSD}{r^2_B}$ for ${\omega}_c$=1.2, ${\gamma}=100$; (c)  $\frac{C(t)}{C(t_0)}$ for ${\omega}_c$=10, ${\gamma}=2$; (d) $\frac{C(t)}{C(t_0)}$ for ${\omega}_c$=1.2, ${\gamma}=100$  in log-log scale; (e) $\frac{C_{p-v}(t)}{C_{p-v}(t_0)}$ for ${\omega}_c$=10, ${\gamma}=2$; (f)$\frac{C_{p-v}(t)}{C_{p-v}(t_0)}$ for ${\omega}_c$=1.2, ${\gamma}=100$  in log-log scale; (g) $\frac{C_v(t)}{C_v(t_0)}$ for ${\omega}_c$=10, ${\gamma}=2$; (h) $\frac{C_v(t)}{C_v(t_0)}$ for ${\omega}_c$=1.2, ${\gamma}=100$  in log-log scale (The scaling time $t_0=\frac{1}{\Omega_{th}}$). [The orange dashed lines in the plots represent the long time exponential behaviours of the correlation functions.]}
\end{figure}
 \end{widetext}

  In Fig.(1a), we notice that the Position Response Function in the under-damped regime initially shows an oscillatory behaviour starting from a finite value and then settles to zero with increasing time. On the other hand, the Response Function in the over-damped regime exhibits an exponential decay from $\frac{{\gamma}}{ m_r {\omega}_0^2}$ at t=0. As discussed in the previous section, this kind of behaviour resembles the Velocity Response Function, obtained as the derivative of the Position Response Function for coupling through position coordinate. The sharp contrast stems from the ${\omega}^3$ dependence of force correlations  (see Eq.(\ref{forcecorrelationA})) in the case of a bath coupled via momentum variables, which is in contrast to the case 
 of a bath coupled via position variables, where the force correlation varies linearly with ${\omega}$. This kind of a behaviour is expected in the case of a bath 
 coupled via momentum variables since the position and momentum are conjugate variables connected by a time derivative.\\
\hspace*{0.2cm}  One noticeable distinctive feature of the plots of the MSD versus $t$ is the fact that the MSD settles to a constant value (2C(0)) at long times over the entire range of temperatures. This feature 
 survives even when one removes the harmonic confinement. This is strikingly different from the case of a bath with a position coordinate coupling, where the MSD exhibits a linear behaviour with time in the absence of a harmonic potential (i.e. for ${\omega}_0 = 0$) and settles to a constant value in the presence of a harmonic confinement (i.e. for a nonzero ${\omega}_0$) \cite{Urbashione}. Similar to the Position Response Function, this behaviour of the MSD is attributed to the decay of position autocorrelation function (C(t)) to zero, with increasing time, due to the fact that the force correlation (see Eq.(\ref{forcecorrelationA})) goes as ${\omega}^3$. \\
 \hspace*{0.2cm} We have also plotted the correlation functions at finite and T$\rightarrow0$ temperatures and analyzed the long time behaviours of the correlation functions, similar to our previously reported work on the quantum Brownian motion of a particle coupled to a bath via a position coordinate coupling \cite{surakatwo}.\\ At finite temperatures, the correlation functions display an exponential decay at long times as discussed in the previous section. However, at $T\rightarrow0$, we notice a transition from a power law to an exponential behaviour around a time scale set by $\frac{\hbar}{2 \pi k_B T }$ \cite{graberttwo,surakatwo}. This can be clearly seen from the above plots, where, the orange dashed lines in (Figs.(1c)-(1h)) represent the long time exponential behaviour at finite temperatures, whilst the orange dashed curves in (Figs.(2c)-(2h)) represent the power law tails at long times for $T\rightarrow0$. It can be noticed that the t$^{-4}$,t$^{-5}$,t$^{-6}$ power law decays of the position autocorrelation function (C(t)), position-velocity correlation function (C$_{p-v}$(t)) and velocity autocorrelation function (C$_v$(t)) respectively are sharper compared to the case of a position coordinate coupling (t$^{-2}$ for C(t), t$^{-3}$ for C$_{p-v}$(t), t$^{-4}$ for C$_v$(t)) \cite{surakatwo}. Hence, it can be inferred from the behaviours of the MSD and the correlation functions, that the effect of harmonic confinement is more pronounced for a particle coupled to the heat bath via momentum coordinate coupling. It may also be noted that, the t$^{-4}$ decay of the position autocorrelation function in the case of a coupling to the bath via a momentum coordinate is exactly same as the t$^{-4}$ power law decay of the velocity autocorrelation function in the case of a coupling to the bath via a position coordinate
 coupling.
 
\section{Conclusion}
 We have derived the mean square displacement, position response function and the position autocorrelation, position-velocity correlation and velocity autocorrelation functions of a charged harmonically confined Quantum Brownian particle in the presence of a magnetic field and attached to a heat bath via a momentum coordinate coupling. There are quite a few 
 studies on the case of a momentum coordinate coupling to a bath of a Quantum Brownian particle \cite{Caldeira,Leggett,Cuccoli,Bai,Bao,Bao2,Ankerhold,Malaymomentum}. Nevertheless, the study of a system coupled to a heat bath via a momentum coordinate coupling still remains a relatively unexplored area in the context of Quantum Brownian motion, in particular 
 in the context of a charged harmonically confined quantum Brownian particle coupled to a heat bath. In \cite{Malaymomentum}
 the Quantum Langevin Equation for a charged harmonic oscillator coupled to a heat bath via a momentum coordinate coupling and in the presence of a magnetic field has been derived. We have, however, gone far beyond that and derived the MSD, the position response function and the correlation functions and analyzed their long time behaviours at finite and near zero temperatures. \\
 \hspace*{0.2cm} Here, we notice two distinctive features in the behaviours of the MSD and position response function (R(t)), which can be seen in Figs.(1a,b), (2a,b) and (3a,b). The MSD settles down to a constant value at long times, which persists even in the absence of a confining potential, whereas, R(t) shows an exponential decay, starting from a finite value at t=0. These behaviours of the MSD and R(t) are in sharp contrast to the case of a particle coupled to a bath via a position coupling \cite{surakaone} and the origin of these anomalous behaviours can be traced to the ${\omega}^3$ dependence of the force correlation functions as given in Eq.(\ref{forcecorrelationA}). Moreover, it can be surmised that the position response function in the case of a coupling via a momentum coordinate behaves as the velocity response function for the conventional case of a particle coupled to a bath via a position coordinate coupling, as position and momentum are conjugate variables connected by a derivative. \\
  \hspace*{0.2cm} Furthermore, the formalism discussed here can be applied to other bath models and for generalized forms of couplings, for instance for a coupling strength varying as ${\omega}^\delta$ at low frequency for various values of $\delta$ and at all temperatures \cite{Aslangul}. The change in the low frequency behaviours of the coupling strength is supposed to change the low frequency correlation function and the long time behaviours as we notice the anomalous behaviours of the MSD and R(t), due to the ${\omega}^3$ dependence of the force correlation in the Fourier domain. This type of anomalous coupling can be realized practically, for example, when one considers the effect of black body electromagnetic field in Josephson junction \cite{Cuccoli,Sols}.\\
 \hspace*{0.2cm} We expect the results of our study to induce the experimentalists to test them in a state of the art 
  cold atom laboratory, along the lines of the experiments carried out for anomalous diffusion of cold atoms \cite{posvelprl,Klappauf,Sagi,Solomon}.

 \section*{Appendix}
 The position autocorrelation function is given by:
 \begin{align}
 C(t)=C_1(t)+C_2(t)
 \end{align}
 where,
\begin{align}
&C_1(t)=(- i)\sum_{j=1}^4 \left( Res[C({\omega})exp(-i{\omega} t), p_j] \right) \\
&C_2(t)=  \left(-i \hbar {\omega}_{th} \right) \sum_{n=1}^\infty R''(-i n \pi {\omega}_{th}) exp(-n \pi {\omega}_{th}t) 
 \end{align}
 Res[C(${\omega}$)$exp(-i{\omega} t)$, p$_j$] is the residue of C(${\omega}$)$exp(-i{\omega} t)$ corresponding to pole at p$_j$
 \begin{widetext}
 \begin{align}
    I_1&= Res[C({\omega})exp(-i{\omega} t), p_1] \notag \\
    & =\left[\frac{e^{-\frac{it\left(-m ({\omega}_c+i{\gamma}) +\sqrt{4 m_r^2 {\omega}_0^2+m^2({\omega}_c+ i {\gamma})^2}\right)}{2 m_r}}\left(2 i m_r^2{\omega}_0^2+m \left(\sqrt{4 m_r^2 {\omega}_0^2 +m^2({\omega}_c+i {\gamma})^2}-m({\omega}_c+i {\gamma})\right)(-i {\omega} _c + {\gamma})\right)\left(\frac{{\gamma} m_r}{ {\omega}_0^2}  \right) \hbar }{8 m m_r^2 {\gamma} \sqrt{4 m_r^2 {\omega}_0^2+m^2( {\omega}_c+ i {\gamma})^2}} \right] \times \notag \\
     &cot h\left(\frac{\sqrt{4m_r^2 {\omega}_0^2 +m^2({\omega}_c+ i {\gamma})^2-m ({\omega}_c+ i {\gamma})}}{2 m_r {\omega}} \right)
 \end{align}\\

 \begin{align}
    I_2&= Res[C({\omega})exp(-i{\omega} t), p_2] \notag \\
    & =\left[\frac{ie^{\frac{it\left(m ({\omega}_c+i{\gamma}) +\sqrt{4 m_r^2 {\omega}_0^2+m^2( {\omega}_c+ i {\gamma})^2}\right)}{2 m_r}}\left(2 m_r^2 {\omega}_0^2+m \left(\sqrt{4 m_r^2 {\omega}_0^2 +m^2({\omega}_c+i {\gamma})^2}+m({\omega}_c+i {\gamma})\right)( {\omega} _c + i {\gamma})\right)\left(\frac{{\gamma} m_r}{ {\omega}_0^2}  \right) \hbar }{8 m m_r^2 {\gamma} \sqrt{4 m_r^2 {\omega}_0^2+m^2( {\omega}_c+ i  {\gamma})^2}} \right] \times \notag \\
     &cot h\left(\frac{\sqrt{4m_r^2 {\omega}_0^2 +m^2({\omega}_c+ i {\gamma})^2+m ({\omega}_c+ i {\gamma})}}{2 m_r {\omega}} \right)
 \end{align}\\
 
 \begin{align}
    I_3&= Res[C({\omega})exp(-i{\omega} t), p_3] \notag \\
    & =\left[\frac{e^{-\frac{it\left(m ({\omega}_c-i{\gamma}) +\sqrt{4 m_r^2 {\omega}_0^2+m^2( {\omega}_c- i {\gamma})^2}\right)}{2 m_r}}\left(2 i m_r^2 {\omega}_0^2+m \left(\sqrt{4 m_r^2 {\omega}_0^2 +m^2({\omega}_c-i {\gamma})^2}+m({\omega}_c-i {\gamma})\right)( i {\omega} _c + {\gamma})\right)\left(\frac{{\gamma} m_r}{ {\omega}_0^2}  \right) \hbar }{8 m m_r^2  {\gamma} \sqrt{4 m_r^2 {\omega}_0^2+m^2( {\omega}_c- i  {\gamma})^2}} \right] \times \notag \\
     &cot h\left(\frac{\sqrt{4m_r^2 {\omega}_0^2 +m^2({\omega}_c- i {\gamma})^2+m ({\omega}_c- i {\gamma})}}{2 m_r {\omega}} \right)
 \end{align} \\
 
 \begin{align}
    I_4&= Res[C({\omega})exp(-i{\omega} t), p_4] \notag \\
    & =\left[\frac{e^{\frac{it\left(m (-{\omega}_c+i{\gamma}) +\sqrt{4 m_r^2 {\omega}_0^2+m^2( {\omega}_c- i {\gamma})^2}\right)}{2 m_r}}\left(2 i m_r^2 {\omega}_0^2+m \left(-\sqrt{4 m_r^2 {\omega}_0^2 +m^2({\omega}_c-i {\gamma})^2}+m({\omega}_c-i {\gamma})\right)( i {\omega} _c + {\gamma})\right)\left(\frac{{\gamma} m_r}{ {\omega}_0^2}  \right) \hbar }{8 m m_r^2 {\gamma} \sqrt{4 m_r^2 {\omega}_0^2+m^2( {\omega}_c- i  {\gamma})^2}} \right] \times \notag \\
     &cot h\left(\frac{\sqrt{4m_r^2 {\omega}_0^2 +m^2({\omega}_c- i {\gamma})^2+m (-{\omega}_c+i {\gamma})}}{2 m_r {\omega}} \right)
 \end{align}
 
 Therefore, the position autocorrelation function is given by:

  \begin{align}
  C(t)&=(- i)\left(I_1+I_2+I_3+I_4 \right)+ \left(-i \hbar {\omega}_{th} \right) \sum_{n=1}^\infty R''(-i n \pi {\omega}_{th}) exp(-n \pi {\omega}_{th}t) \notag \\
    & =(- i)\left(I_1+I_2+I_3+I_4 \right)+\frac{\hbar \pi^2 {\omega}_{th}^4 {\gamma}}{m m_r {\omega}_0^6} \frac{e^{\pi {\omega}_{th}t}(1+4e^{\pi {\omega}_{th}t}+e^{2\pi {\omega}_{th}t})}{(e^{\pi {\omega}_{th}t}-1)^4}- \frac{\hbar \pi^4 {\omega}_{th}^6 {\gamma}(2 m_r^1 {\omega}_0^2-m^2({\gamma}^2-3 {\omega}_c^2))}{m m_r^3 {\omega}_0^{10}}\times  \notag\\
    &\frac{e^{\pi {\omega}_{th}t}(1+26e^{\pi {\omega}_{th}t}+66e^{2\pi {\omega}_{th}t}+26e^{3\pi {\omega}_{th}t}+e^{4\pi {\omega}_{th}t})}{(e^{\pi {\omega}_{th}t}-1)^6} +.......
 \end{align}
 \end{widetext}




\begin{thebibliography}{33}%
\makeatletter
\providecommand \@ifxundefined [1]{%
 \@ifx{#1\undefined}
}%
\providecommand \@ifnum [1]{%
 \ifnum #1\expandafter \@firstoftwo
 \else \expandafter \@secondoftwo
 \fi
}%
\providecommand \@ifx [1]{%
 \ifx #1\expandafter \@firstoftwo
 \else \expandafter \@secondoftwo
 \fi
}%
\providecommand \natexlab [1]{#1}%
\providecommand \enquote  [1]{``#1''}%
\providecommand \bibnamefont  [1]{#1}%
\providecommand \bibfnamefont [1]{#1}%
\providecommand \citenamefont [1]{#1}%
\providecommand \href@noop [0]{\@secondoftwo}%
\providecommand \href [0]{\begingroup \@sanitize@url \@href}%
\providecommand \@href[1]{\@@startlink{#1}\@@href}%
\providecommand \@@href[1]{\endgroup#1\@@endlink}%
\providecommand \@sanitize@url [0]{\catcode `\\12\catcode `\$12\catcode
  `\&12\catcode `\#12\catcode `\^12\catcode `\_12\catcode `\%12\relax}%
\providecommand \@@startlink[1]{}%
\providecommand \@@endlink[0]{}%
\providecommand \url  [0]{\begingroup\@sanitize@url \@url }%
\providecommand \@url [1]{\endgroup\@href {#1}{\urlprefix }}%
\providecommand \urlprefix  [0]{URL }%
\providecommand \Eprint [0]{\href }%
\providecommand \doibase [0]{https://doi.org/}%
\providecommand \selectlanguage [0]{\@gobble}%
\providecommand \bibinfo  [0]{\@secondoftwo}%
\providecommand \bibfield  [0]{\@secondoftwo}%
\providecommand \translation [1]{[#1]}%
\providecommand \BibitemOpen [0]{}%
\providecommand \bibitemStop [0]{}%
\providecommand \bibitemNoStop [0]{.\EOS\space}%
\providecommand \EOS [0]{\spacefactor3000\relax}%
\providecommand \BibitemShut  [1]{\csname bibitem#1\endcsname}%
\let\auto@bib@innerbib\@empty
\bibitem [{\citenamefont {Grabert}\ \emph {et~al.}(1984)\citenamefont
  {Grabert}, \citenamefont {Weiss},\ and\ \citenamefont
  {Talkner}}]{grabertone}%
  \BibitemOpen
  \bibfield  {author} {\bibinfo {author} {\bibfnamefont {H.}~\bibnamefont
  {Grabert}}, \bibinfo {author} {\bibfnamefont {U.}~\bibnamefont {Weiss}},\
  and\ \bibinfo {author} {\bibfnamefont {P.}~\bibnamefont {Talkner}},\
  }\bibfield  {title} {\bibinfo {title} {Quantum theory of the damped harmonic
  oscillator},\ }\href {https://doi.org/10.1007/BF01307505} {\bibfield
  {journal} {\bibinfo  {journal} {Z. Physik B - Condensed Matter}\ }\textbf
  {\bibinfo {volume} {55}},\ \bibinfo {pages} {87} (\bibinfo {year}
  {1984})}\BibitemShut {NoStop}%
\bibitem [{\citenamefont {Jung}\ \emph {et~al.}(1985)\citenamefont {Jung},
  \citenamefont {Ingold},\ and\ \citenamefont {Grabert}}]{graberttwo}%
  \BibitemOpen
  \bibfield  {author} {\bibinfo {author} {\bibfnamefont {R.}~\bibnamefont
  {Jung}}, \bibinfo {author} {\bibfnamefont {G.}~\bibnamefont {Ingold}},\ and\
  \bibinfo {author} {\bibfnamefont {H.}~\bibnamefont {Grabert}},\ }\bibfield
  {title} {\bibinfo {title} {Long-time tails in quantum brownian motion},\
  }\href {https://doi.org/10.1103/PhysRevA.32.2510} {\bibfield  {journal}
  {\bibinfo  {journal} {Phys. Rev. A}\ }\textbf {\bibinfo {volume} {32}},\
  \bibinfo {pages} {2510} (\bibinfo {year} {1985})}\BibitemShut {NoStop}%
\bibitem [{\citenamefont {S.Sinha}(1997)}]{decosup}%
  \BibitemOpen
  \bibfield  {author} {\bibinfo {author} {\bibnamefont {S.Sinha}},\ }\bibfield
  {title} {\bibinfo {title} {Decoherence at absolute zero},\ }\href
  {https://doi.org/10.1016/S0375-9601(97)00098-4} {\bibfield  {journal}
  {\bibinfo  {journal} {Physics Letters A}\ }\textbf {\bibinfo {volume}
  {228}},\ \bibinfo {pages} {1} (\bibinfo {year} {1997})}\BibitemShut {NoStop}%
\bibitem [{\citenamefont {Aslangul}\ \emph {et~al.}(1987)\citenamefont
  {Aslangul}, \citenamefont {Pottier},\ and\ \citenamefont
  {Saint-James}}]{Aslangul}%
  \BibitemOpen
  \bibfield  {author} {\bibinfo {author} {\bibfnamefont {C.}~\bibnamefont
  {Aslangul}}, \bibinfo {author} {\bibfnamefont {N.}~\bibnamefont {Pottier}},\
  and\ \bibinfo {author} {\bibfnamefont {D.}~\bibnamefont {Saint-James}},\
  }\bibfield  {title} {\bibinfo {title} {Quantum dynamics of a damped free
  particle},\ }\href {https://doi.org/10.1051/jphys:0198700480110187100}
  {\bibfield  {journal} {\bibinfo  {journal} {J. Phys. France}\ }\textbf
  {\bibinfo {volume} {48}},\ \bibinfo {pages} {1871} (\bibinfo {year}
  {1987})}\BibitemShut {NoStop}%
\bibitem [{\citenamefont {Ford}\ \emph {et~al.}(1988)\citenamefont {Ford},
  \citenamefont {Lewis},\ and\ \citenamefont {O'Connell}}]{ford}%
  \BibitemOpen
  \bibfield  {author} {\bibinfo {author} {\bibfnamefont {G.~W.}\ \bibnamefont
  {Ford}}, \bibinfo {author} {\bibfnamefont {J.~T.}\ \bibnamefont {Lewis}},\
  and\ \bibinfo {author} {\bibfnamefont {R.~F.}\ \bibnamefont {O'Connell}},\
  }\bibfield  {title} {\bibinfo {title} {Quantum langevin equation},\ }\href
  {https://doi.org/10.1103/PhysRevA.37.4419} {\bibfield  {journal} {\bibinfo
  {journal} {Phys. Rev. A}\ }\textbf {\bibinfo {volume} {37}},\ \bibinfo
  {pages} {4419} (\bibinfo {year} {1988})}\BibitemShut {NoStop}%
\bibitem [{\citenamefont {Satpathi}\ and\ \citenamefont
  {Sinha}(2018)}]{Urbashione}%
  \BibitemOpen
  \bibfield  {author} {\bibinfo {author} {\bibfnamefont {U.}~\bibnamefont
  {Satpathi}}\ and\ \bibinfo {author} {\bibfnamefont {S.}~\bibnamefont
  {Sinha}},\ }\bibfield  {title} {\bibinfo {title} {Quantum {B}rownian motion
  in a magnetic field: Transition from monotonic to oscillatory behaviour},\
  }\href {https://doi.org/https://doi.org/10.1016/j.physa.2018.04.085}
  {\bibfield  {journal} {\bibinfo  {journal} {Physica A: Statistical Mechanics
  and its Applications}\ }\textbf {\bibinfo {volume} {506}},\ \bibinfo {pages}
  {692} (\bibinfo {year} {2018})}\BibitemShut {NoStop}%
\bibitem [{\citenamefont {Bhattacharjee}\ \emph
  {et~al.}(2022{\natexlab{a}})\citenamefont {Bhattacharjee}, \citenamefont
  {Satpathi},\ and\ \citenamefont {Sinha}}]{surakaone}%
  \BibitemOpen
  \bibfield  {author} {\bibinfo {author} {\bibfnamefont {S.}~\bibnamefont
  {Bhattacharjee}}, \bibinfo {author} {\bibfnamefont {U.}~\bibnamefont
  {Satpathi}},\ and\ \bibinfo {author} {\bibfnamefont {S.}~\bibnamefont
  {Sinha}},\ }\bibfield  {title} {\bibinfo {title} {Quantum langevin dynamics
  of a charged particle in a magnetic field: Response function,
  position–velocity and velocity autocorrelation functions},\ }\href
  {https://doi.org/10.1007/s12043-022-02295-1} {\bibfield  {journal} {\bibinfo
  {journal} {Pramana}\ }\textbf {\bibinfo {volume} {96}},\ \bibinfo {pages}
  {53} (\bibinfo {year} {2022}{\natexlab{a}})}\BibitemShut {NoStop}%
\bibitem [{\citenamefont {Li}\ \emph {et~al.}(1990)\citenamefont {Li},
  \citenamefont {Ford},\ and\ \citenamefont {O'Connell}}]{Li}%
  \BibitemOpen
  \bibfield  {author} {\bibinfo {author} {\bibfnamefont {X.~L.}\ \bibnamefont
  {Li}}, \bibinfo {author} {\bibfnamefont {G.~W.}\ \bibnamefont {Ford}},\ and\
  \bibinfo {author} {\bibfnamefont {R.~F.}\ \bibnamefont {O'Connell}},\
  }\bibfield  {title} {\bibinfo {title} {Magnetic-field effects on the motion
  of a charged particle in a heat bath},\ }\href
  {https://doi.org/10.1103/PhysRevA.41.5287} {\bibfield  {journal} {\bibinfo
  {journal} {Phys. Rev. A}\ }\textbf {\bibinfo {volume} {41}},\ \bibinfo
  {pages} {5287} (\bibinfo {year} {1990})}\BibitemShut {NoStop}%
\bibitem [{\citenamefont {Fa}(2009)}]{fa}%
  \BibitemOpen
  \bibfield  {author} {\bibinfo {author} {\bibfnamefont {K.}~\bibnamefont
  {Fa}},\ }\bibfield  {title} {\bibinfo {title} {Anomalous diffusion in a
  generalized langevin equation},\ }\href {https://doi.org/10.1063/1.3187218}
  {\bibfield  {journal} {\bibinfo  {journal} {Journal of Mathematical Physics}\
  }\textbf {\bibinfo {volume} {50}},\ \bibinfo {pages} {083301} (\bibinfo
  {year} {2009})}\BibitemShut {NoStop}%
\bibitem [{\citenamefont {Mckinley}\ and\ \citenamefont
  {Nguyen}(2018)}]{Mckinley}%
  \BibitemOpen
  \bibfield  {author} {\bibinfo {author} {\bibfnamefont {S.}~\bibnamefont
  {Mckinley}}\ and\ \bibinfo {author} {\bibfnamefont {H.}~\bibnamefont
  {Nguyen}},\ }\bibfield  {title} {\bibinfo {title} {Anomalous diffusion and
  the generalized langevin equation},\ }\href
  {https://doi.org/10.1137/17M115517X} {\bibfield  {journal} {\bibinfo
  {journal} {Siam.J.Math.Anal}\ }\textbf {\bibinfo {volume} {50}},\ \bibinfo
  {pages} {5119–5160} (\bibinfo {year} {2018})}\BibitemShut {NoStop}%
\bibitem [{\citenamefont {Didier}\ and\ \citenamefont {Nguyen}(2022)}]{Didier}%
  \BibitemOpen
  \bibfield  {author} {\bibinfo {author} {\bibfnamefont {G.}~\bibnamefont
  {Didier}}\ and\ \bibinfo {author} {\bibfnamefont {H.}~\bibnamefont
  {Nguyen}},\ }\bibfield  {title} {\bibinfo {title} {The generalized langevin
  equation in harmonic potentials: Anomalous diffusion and equipartition of
  energy},\ }\href {https://doi.org/10.1007/s00220-022-04378-x} {\bibfield
  {journal} {\bibinfo  {journal} {Communications in Mathematical Physics}\ }
  (\bibinfo {year} {2022})}\BibitemShut {NoStop}%
\bibitem [{\citenamefont {Lis\'y}\ and\ \citenamefont
  {T\'othov\'a}(2019)}]{Lisy}%
  \BibitemOpen
  \bibfield  {author} {\bibinfo {author} {\bibfnamefont {V.}~\bibnamefont
  {Lis\'y}}\ and\ \bibinfo {author} {\bibfnamefont {J.}~\bibnamefont
  {T\'othov\'a}},\ }\bibfield  {title} {\bibinfo {title} {Generalized langevin
  equation and the fluctuation-dissipation theorem for particle-bath systems in
  a harmonic field},\ }\href
  {https://doi.org/https://doi.org/10.1016/j.rinp.2019.01.003} {\bibfield
  {journal} {\bibinfo  {journal} {Results in Physics}\ }\textbf {\bibinfo
  {volume} {12}},\ \bibinfo {pages} {1212} (\bibinfo {year}
  {2019})}\BibitemShut {NoStop}%
\bibitem [{\citenamefont {T\'othov\'a}\ and\ \citenamefont
  {Lis\'y}(2021)}]{Tothova}%
  \BibitemOpen
  \bibfield  {author} {\bibinfo {author} {\bibfnamefont {J.}~\bibnamefont
  {T\'othov\'a}}\ and\ \bibinfo {author} {\bibfnamefont {V.}~\bibnamefont
  {Lis\'y}},\ }\bibfield  {title} {\bibinfo {title} {Brownian motion in a bath
  affected by an external harmonic potential},\ }\href
  {https://doi.org/https://doi.org/10.1016/j.physleta.2021.127220} {\bibfield
  {journal} {\bibinfo  {journal} {Physics Letters A}\ }\textbf {\bibinfo
  {volume} {395}},\ \bibinfo {pages} {127220} (\bibinfo {year}
  {2021})}\BibitemShut {NoStop}%
\bibitem [{\citenamefont {Berner}\ \emph {et~al.}(2018)\citenamefont {Berner},
  \citenamefont {M\"uller}, \citenamefont {Gomez-Solano}, \citenamefont
  {Kr\"uger},\ and\ \citenamefont {Bechinger}}]{Berner}%
  \BibitemOpen
  \bibfield  {author} {\bibinfo {author} {\bibfnamefont {J.}~\bibnamefont
  {Berner}}, \bibinfo {author} {\bibfnamefont {B.}~\bibnamefont {M\"uller}},
  \bibinfo {author} {\bibfnamefont {J.}~\bibnamefont {Gomez-Solano}}, \bibinfo
  {author} {\bibfnamefont {M.}~\bibnamefont {Kr\"uger}},\ and\ \bibinfo
  {author} {\bibfnamefont {C.}~\bibnamefont {Bechinger}},\ }\bibfield  {title}
  {\bibinfo {title} {Oscillating modes of driven colloids in overdamped
  systems},\ }\href
  {https://doi.org/https://doi.org/10.1038/s41467-018-03345-2} {\bibfield
  {journal} {\bibinfo  {journal} {Nature Communications}\ }\textbf {\bibinfo
  {volume} {9}},\ \bibinfo {pages} {999} (\bibinfo {year} {2018})}\BibitemShut
  {NoStop}%
\bibitem [{\citenamefont {Daldrop}\ \emph {et~al.}(2017)\citenamefont
  {Daldrop}, \citenamefont {Kowalik},\ and\ \citenamefont {Netz}}]{Daldrop}%
  \BibitemOpen
  \bibfield  {author} {\bibinfo {author} {\bibfnamefont {J.~O.}\ \bibnamefont
  {Daldrop}}, \bibinfo {author} {\bibfnamefont {B.~G.}\ \bibnamefont
  {Kowalik}},\ and\ \bibinfo {author} {\bibfnamefont {R.~R.}\ \bibnamefont
  {Netz}},\ }\bibfield  {title} {\bibinfo {title} {External potential modifies
  friction of molecular solutes in water},\ }\href
  {https://doi.org/10.1103/PhysRevX.7.041065} {\bibfield  {journal} {\bibinfo
  {journal} {Phys. Rev. X}\ }\textbf {\bibinfo {volume} {7}},\ \bibinfo {pages}
  {041065} (\bibinfo {year} {2017})}\BibitemShut {NoStop}%
\bibitem [{\citenamefont {Caldeira}\ and\ \citenamefont
  {Leggett}(1983)}]{Caldeira}%
  \BibitemOpen
  \bibfield  {author} {\bibinfo {author} {\bibfnamefont {A.}~\bibnamefont
  {Caldeira}}\ and\ \bibinfo {author} {\bibfnamefont {A.}~\bibnamefont
  {Leggett}},\ }\bibfield  {title} {\bibinfo {title} {Path integral approach to
  quantum brownian motion},\ }\href
  {https://doi.org/https://doi.org/10.1016/0378-4371(83)90013-4} {\bibfield
  {journal} {\bibinfo  {journal} {Physica A: Statistical Mechanics and its
  Applications}\ }\textbf {\bibinfo {volume} {121}},\ \bibinfo {pages} {587}
  (\bibinfo {year} {1983})}\BibitemShut {NoStop}%
\bibitem [{\citenamefont {Leggett}(1984)}]{Leggett}%
  \BibitemOpen
  \bibfield  {author} {\bibinfo {author} {\bibfnamefont {A.~J.}\ \bibnamefont
  {Leggett}},\ }\bibfield  {title} {\bibinfo {title} {Quantum tunneling in the
  presence of an arbitrary linear dissipation mechanism},\ }\href
  {https://doi.org/10.1103/PhysRevB.30.1208} {\bibfield  {journal} {\bibinfo
  {journal} {Phys. Rev. B}\ }\textbf {\bibinfo {volume} {30}},\ \bibinfo
  {pages} {1208} (\bibinfo {year} {1984})}\BibitemShut {NoStop}%
\bibitem [{\citenamefont {Cuccoli}\ \emph {et~al.}(2001)\citenamefont
  {Cuccoli}, \citenamefont {Fubini}, \citenamefont {Tognetti},\ and\
  \citenamefont {Vaia}}]{Cuccoli}%
  \BibitemOpen
  \bibfield  {author} {\bibinfo {author} {\bibfnamefont {A.}~\bibnamefont
  {Cuccoli}}, \bibinfo {author} {\bibfnamefont {A.}~\bibnamefont {Fubini}},
  \bibinfo {author} {\bibfnamefont {V.}~\bibnamefont {Tognetti}},\ and\
  \bibinfo {author} {\bibfnamefont {R.}~\bibnamefont {Vaia}},\ }\bibfield
  {title} {\bibinfo {title} {Quantum thermodynamics of systems with anomalous
  dissipative coupling},\ }\href {https://doi.org/10.1103/PhysRevE.64.066124}
  {\bibfield  {journal} {\bibinfo  {journal} {Phys. Rev. E}\ }\textbf {\bibinfo
  {volume} {64}},\ \bibinfo {pages} {066124} (\bibinfo {year}
  {2001})}\BibitemShut {NoStop}%
\bibitem [{\citenamefont {Bai}\ \emph {et~al.}(2005)\citenamefont {Bai},
  \citenamefont {Bao},\ and\ \citenamefont {Song}}]{Bai}%
  \BibitemOpen
  \bibfield  {author} {\bibinfo {author} {\bibfnamefont {Z.-W.}\ \bibnamefont
  {Bai}}, \bibinfo {author} {\bibfnamefont {J.-D.}\ \bibnamefont {Bao}},\ and\
  \bibinfo {author} {\bibfnamefont {Y.-L.}\ \bibnamefont {Song}},\ }\bibfield
  {title} {\bibinfo {title} {Classical and quantum diffusion in the presence of
  velocity-dependent coupling},\ }\href
  {https://doi.org/10.1103/PhysRevE.72.061105} {\bibfield  {journal} {\bibinfo
  {journal} {Phys. Rev. E}\ }\textbf {\bibinfo {volume} {72}},\ \bibinfo
  {pages} {061105} (\bibinfo {year} {2005})}\BibitemShut {NoStop}%
\bibitem [{\citenamefont {Bao}\ and\ \citenamefont {Zhuo}(2005)}]{Bao}%
  \BibitemOpen
  \bibfield  {author} {\bibinfo {author} {\bibfnamefont {J.-D.}\ \bibnamefont
  {Bao}}\ and\ \bibinfo {author} {\bibfnamefont {Y.-Z.}\ \bibnamefont {Zhuo}},\
  }\bibfield  {title} {\bibinfo {title} {Anomalous dissipation: Strong
  non-markovian effect and its dynamical origin},\ }\href
  {https://doi.org/10.1103/PhysRevE.71.010102} {\bibfield  {journal} {\bibinfo
  {journal} {Phys. Rev. E}\ }\textbf {\bibinfo {volume} {71}},\ \bibinfo
  {pages} {010102} (\bibinfo {year} {2005})}\BibitemShut {NoStop}%
\bibitem [{\citenamefont {Bao}\ \emph {et~al.}(2006)\citenamefont {Bao},
  \citenamefont {Zhuo}, \citenamefont {Oliveira},\ and\ \citenamefont
  {H\"anggi}}]{Bao2}%
  \BibitemOpen
  \bibfield  {author} {\bibinfo {author} {\bibfnamefont {J.-D.}\ \bibnamefont
  {Bao}}, \bibinfo {author} {\bibfnamefont {Y.-Z.}\ \bibnamefont {Zhuo}},
  \bibinfo {author} {\bibfnamefont {F.~A.}\ \bibnamefont {Oliveira}},\ and\
  \bibinfo {author} {\bibfnamefont {P.}~\bibnamefont {H\"anggi}},\ }\bibfield
  {title} {\bibinfo {title} {Intermediate dynamics between newton and
  langevin},\ }\href {https://doi.org/10.1103/PhysRevE.74.061111} {\bibfield
  {journal} {\bibinfo  {journal} {Phys. Rev. E}\ }\textbf {\bibinfo {volume}
  {74}},\ \bibinfo {pages} {061111} (\bibinfo {year} {2006})}\BibitemShut
  {NoStop}%
\bibitem [{\citenamefont {Ankerhold}\ and\ \citenamefont
  {Pollak}(2007)}]{Ankerhold}%
  \BibitemOpen
  \bibfield  {author} {\bibinfo {author} {\bibfnamefont {J.}~\bibnamefont
  {Ankerhold}}\ and\ \bibinfo {author} {\bibfnamefont {E.}~\bibnamefont
  {Pollak}},\ }\bibfield  {title} {\bibinfo {title} {Dissipation can enhance
  quantum effects},\ }\href {https://doi.org/10.1103/PhysRevE.75.041103}
  {\bibfield  {journal} {\bibinfo  {journal} {Phys. Rev. E}\ }\textbf {\bibinfo
  {volume} {75}},\ \bibinfo {pages} {041103} (\bibinfo {year}
  {2007})}\BibitemShut {NoStop}%
\bibitem [{\citenamefont {Gupta}\ and\ \citenamefont
  {Bandyopadhyay}(2011)}]{Malaymomentum}%
  \BibitemOpen
  \bibfield  {author} {\bibinfo {author} {\bibfnamefont {S.}~\bibnamefont
  {Gupta}}\ and\ \bibinfo {author} {\bibfnamefont {M.}~\bibnamefont
  {Bandyopadhyay}},\ }\bibfield  {title} {\bibinfo {title} {Quantum langevin
  equation of a charged oscillator in a magnetic field and coupled to a heat
  bath through momentum variables},\ }\href
  {https://doi.org/10.1103/PhysRevE.84.041133} {\bibfield  {journal} {\bibinfo
  {journal} {Phys. Rev. E}\ }\textbf {\bibinfo {volume} {84}},\ \bibinfo
  {pages} {041133} (\bibinfo {year} {2011})}\BibitemShut {NoStop}%
\bibitem [{\citenamefont {Bhattacharjee}\ \emph
  {et~al.}(2022{\natexlab{b}})\citenamefont {Bhattacharjee}, \citenamefont
  {Satpathi},\ and\ \citenamefont {Sinha}}]{surakatwo}%
  \BibitemOpen
  \bibfield  {author} {\bibinfo {author} {\bibfnamefont {S.}~\bibnamefont
  {Bhattacharjee}}, \bibinfo {author} {\bibfnamefont {U.}~\bibnamefont
  {Satpathi}},\ and\ \bibinfo {author} {\bibfnamefont {S.}~\bibnamefont
  {Sinha}},\ }\bibfield  {title} {\bibinfo {title} {Long time tails in quantum
  brownian motion of a charged particle in a magnetic field},\ }\href@noop {}
  {\bibfield  {journal} {\bibinfo  {journal} {arXiv: 2201.09712}\ } (\bibinfo
  {year} {2022}{\natexlab{b}})}\BibitemShut {NoStop}%
\bibitem [{\citenamefont {Felderhof}(1978)}]{Felderhof1}%
  \BibitemOpen
  \bibfield  {author} {\bibinfo {author} {\bibfnamefont {B.~U.}\ \bibnamefont
  {Felderhof}},\ }\bibfield  {title} {\bibinfo {title} {On the derivation of
  the fluctuation-dissipation theorem},\ }\href
  {https://doi.org/10.1088/0305-4470/11/5/021} {\bibfield  {journal} {\bibinfo
  {journal} {Journal of Physics A: Mathematical and General}\ }\textbf
  {\bibinfo {volume} {11}},\ \bibinfo {pages} {921} (\bibinfo {year}
  {1978})}\BibitemShut {NoStop}%
\bibitem [{\citenamefont {Weber}(1956)}]{Weber}%
  \BibitemOpen
  \bibfield  {author} {\bibinfo {author} {\bibfnamefont {J.}~\bibnamefont
  {Weber}},\ }\bibfield  {title} {\bibinfo {title} {Fluctuation dissipation
  theorem},\ }\href {https://doi.org/10.1103/PhysRev.101.1620} {\bibfield
  {journal} {\bibinfo  {journal} {Phys. Rev.}\ }\textbf {\bibinfo {volume}
  {101}},\ \bibinfo {pages} {1620} (\bibinfo {year} {1956})}\BibitemShut
  {NoStop}%
\bibitem [{\citenamefont {U.Satpathi}\ and\ \citenamefont
  {S.Sinha}(2019)}]{Urbashitwo}%
  \BibitemOpen
  \bibfield  {author} {\bibinfo {author} {\bibnamefont {U.Satpathi}}\ and\
  \bibinfo {author} {\bibnamefont {S.Sinha}},\ }\bibfield  {title} {\bibinfo
  {title} {Non-equilibrium quantum langevin dynamics of orbital diamagnetic
  moment},\ }\href {https://doi.org/10.1088/1742-5468/ab1dda} {\bibfield
  {journal} {\bibinfo  {journal} {Journal of Statistical Mechanics: Theory and
  Experiment}\ }\textbf {\bibinfo {volume} {2019}},\ \bibinfo {pages} {063106}
  (\bibinfo {year} {2019})}\BibitemShut {NoStop}%
\bibitem [{\citenamefont {Bohren}(2010)}]{bohren2010did}%
  \BibitemOpen
  \bibfield  {author} {\bibinfo {author} {\bibfnamefont {C.~F.}\ \bibnamefont
  {Bohren}},\ }\bibfield  {title} {\bibinfo {title} {What did {K}ramers and
  {K}ronig do and how did they do it?},\ }\href
  {https://doi.org/10.1088/0143-0807/31/3/014} {\bibfield  {journal} {\bibinfo
  {journal} {European Journal of Physics}\ }\textbf {\bibinfo {volume} {31}},\
  \bibinfo {pages} {573} (\bibinfo {year} {2010})}\BibitemShut {NoStop}%
\bibitem [{\citenamefont {Sols}\ and\ \citenamefont {Zapata}(1997)}]{Sols}%
  \BibitemOpen
  \bibfield  {author} {\bibinfo {author} {\bibfnamefont {F.}~\bibnamefont
  {Sols}}\ and\ \bibinfo {author} {\bibfnamefont {I.}~\bibnamefont {Zapata}},\
  }\bibinfo {title} {Effect of qed fluctuations on the dynamics of the
  macroscopic phase},\ in\ \href {https://doi.org/10.1007/978-94-011-5886-2_50}
  {\emph {\bibinfo {booktitle} {New Developments on Fundamental Problems in
  Quantum Physics}}},\ \bibinfo {editor} {edited by\ \bibinfo {editor}
  {\bibfnamefont {M.}~\bibnamefont {Ferrero}}\ and\ \bibinfo {editor}
  {\bibfnamefont {A.}~\bibnamefont {van~der Merwe}}}\ (\bibinfo  {publisher}
  {Springer Netherlands},\ \bibinfo {address} {Dordrecht},\ \bibinfo {year}
  {1997})\ pp.\ \bibinfo {pages} {403--413}\BibitemShut {NoStop}%
\bibitem [{\citenamefont {Afek}\ \emph {et~al.}(2017)\citenamefont {Afek},
  \citenamefont {Coslovsky}, \citenamefont {Courvoisier}, \citenamefont
  {Livneh},\ and\ \citenamefont {Davidson}}]{posvelprl}%
  \BibitemOpen
  \bibfield  {author} {\bibinfo {author} {\bibfnamefont {G.}~\bibnamefont
  {Afek}}, \bibinfo {author} {\bibfnamefont {J.}~\bibnamefont {Coslovsky}},
  \bibinfo {author} {\bibfnamefont {A.}~\bibnamefont {Courvoisier}}, \bibinfo
  {author} {\bibfnamefont {O.}~\bibnamefont {Livneh}},\ and\ \bibinfo {author}
  {\bibfnamefont {N.}~\bibnamefont {Davidson}},\ }\bibfield  {title} {\bibinfo
  {title} {Observing power-law dynamics of position-velocity correlation in
  anomalous diffusion},\ }\href
  {https://doi.org/10.1103/PhysRevLett.119.060602} {\bibfield  {journal}
  {\bibinfo  {journal} {Phys. Rev. Lett.}\ }\textbf {\bibinfo {volume} {119}},\
  \bibinfo {pages} {060602} (\bibinfo {year} {2017})}\BibitemShut {NoStop}%
\bibitem [{\citenamefont {Klappauf}\ \emph {et~al.}(1998)\citenamefont
  {Klappauf}, \citenamefont {Oskay}, \citenamefont {Steck},\ and\ \citenamefont
  {Raizen}}]{Klappauf}%
  \BibitemOpen
  \bibfield  {author} {\bibinfo {author} {\bibfnamefont {B.~G.}\ \bibnamefont
  {Klappauf}}, \bibinfo {author} {\bibfnamefont {W.~H.}\ \bibnamefont {Oskay}},
  \bibinfo {author} {\bibfnamefont {D.~A.}\ \bibnamefont {Steck}},\ and\
  \bibinfo {author} {\bibfnamefont {M.~G.}\ \bibnamefont {Raizen}},\ }\bibfield
   {title} {\bibinfo {title} {Experimental study of quantum dynamics in a
  regime of classical anomalous diffusion},\ }\href
  {https://doi.org/10.1103/PhysRevLett.81.4044} {\bibfield  {journal} {\bibinfo
   {journal} {Phys. Rev. Lett.}\ }\textbf {\bibinfo {volume} {81}},\ \bibinfo
  {pages} {4044} (\bibinfo {year} {1998})}\BibitemShut {NoStop}%
\bibitem [{\citenamefont {Sagi}\ \emph {et~al.}(2012)\citenamefont {Sagi},
  \citenamefont {Brook}, \citenamefont {Almog},\ and\ \citenamefont
  {Davidson}}]{Sagi}%
  \BibitemOpen
  \bibfield  {author} {\bibinfo {author} {\bibfnamefont {Y.}~\bibnamefont
  {Sagi}}, \bibinfo {author} {\bibfnamefont {M.}~\bibnamefont {Brook}},
  \bibinfo {author} {\bibfnamefont {I.}~\bibnamefont {Almog}},\ and\ \bibinfo
  {author} {\bibfnamefont {N.}~\bibnamefont {Davidson}},\ }\bibfield  {title}
  {\bibinfo {title} {Observation of anomalous diffusion and fractional
  self-similarity in one dimension},\ }\href
  {https://doi.org/10.1103/PhysRevLett.108.093002} {\bibfield  {journal}
  {\bibinfo  {journal} {Phys. Rev. Lett.}\ }\textbf {\bibinfo {volume} {108}},\
  \bibinfo {pages} {093002} (\bibinfo {year} {2012})}\BibitemShut {NoStop}%
\bibitem [{\citenamefont {Solomon}\ \emph {et~al.}(1993)\citenamefont
  {Solomon}, \citenamefont {Weeks},\ and\ \citenamefont {Swinney}}]{Solomon}%
  \BibitemOpen
  \bibfield  {author} {\bibinfo {author} {\bibfnamefont {T.~H.}\ \bibnamefont
  {Solomon}}, \bibinfo {author} {\bibfnamefont {E.~R.}\ \bibnamefont {Weeks}},\
  and\ \bibinfo {author} {\bibfnamefont {H.~L.}\ \bibnamefont {Swinney}},\
  }\bibfield  {title} {\bibinfo {title} {Observation of anomalous diffusion and
  l\'evy flights in a two-dimensional rotating flow},\ }\href
  {https://doi.org/10.1103/PhysRevLett.71.3975} {\bibfield  {journal} {\bibinfo
   {journal} {Phys. Rev. Lett.}\ }\textbf {\bibinfo {volume} {71}},\ \bibinfo
  {pages} {3975} (\bibinfo {year} {1993})}\BibitemShut {NoStop}%
\end{thebibliography}
%
 \end{document}